\documentclass[12pt]{article}

\usepackage{amsmath}
\usepackage{amssymb}
\usepackage[dvips]{graphicx}
\usepackage{epsf}
\usepackage{comment}

\usepackage[usenames]{color}

\begin{document}

\begin{titlepage}
\begin{center}

\hfill RIKEN-MP-47 \\

\vspace{1.0cm}
{\Large\bf Natural supersymmetric spectrum \\ 
           in mirage mediation }
\vspace{2.0cm}

{\bf Masaki Asano}$^{(a)}$ and {\bf Tetsutaro Higaki}$^{(b)}$

\vspace{1.0cm}
{\it
$^{(a)}${\it II. Institute for Theoretical Physics, 
University of Hamburg, \\
Luruper Chausse 149, DE-22761 Hamburg, Germany} \\
$^{(b)}$ 
Mathematical Physics Lab., RIKEN Nishina Center, Saitama 351-0198, Japan
}
\vspace{2.0cm}

\abstract{ 
Current results of LHC experiments exclude large area of light new particle 
region, namely natural parameter space, in supersymmetric extension 
models. 
One of the possibilities for achieving the correct electroweak symmetry breaking 
naturally is low scale messenger scenario. Actually, the next-to-minimal 
supersymmetric standard model with TeV scale mirage mediation realizes the natural 
electroweak symmetry breaking with various mass spectra. In this paper, we show  
the possible mass spectrum in the scenario, e.g. degenerate and/or hierarchical 
mass spectrum, and discuss these features. 

}

\end{center}
\end{titlepage}
\setcounter{footnote}{0}

\section{Introduction}
\label{sec: intro}

Models of low energy supersymmetry (SUSY) is one of the leading candidates for 
new physics. The SUSY assures the cancellation of the quadratic divergence 
in Higgs mass squared terms and the model realizes light Higgs boson which is 
consistent with electroweak precision measurements and current results of Higgs 
boson search \cite{Barate:2003sz,TEVNPH:2012ab,Chatrchyan:2012tx,HiggsATLAS}.

Naturalness discussion suggests light SUSY particles 
: If the SUSY 
breaking soft masses are much higher than the electroweak symmetry breaking 
(EWSB) scale, serious fine tuning would be needed to realize the correct EWSB. 
On the other hand, the current results of SUSY searches  
at 
LHC experiments 
have already excluded the gluino and squarks lighter than 1 TeV \cite{Aad:2011qa,Chatrchyan:2011zy} if their masses are nearly equal. Moreover, 
recent results for Higgs boson search would imply 
the
existence of the large 
radiative correction from heavy scalar top (stop).

The low messenger scale scenario is one of the possibilities to explain this 
paradoxical results. In the scenario, there is only the small logarithm in the 
radiative corrections of the Higgs mass parameter. Then, relatively large SUSY 
breaking soft masses can achieve the correct EWSB naturally. 

In this paper, we consider the mirage mediation of the SUSY breaking 
\cite{Choi:2004sx, Choi:2005ge, Choi:2005uz, Endo:2005uy} as the low messenger 
scale scenario. In the mirage mediation scenario, 
SUSY breaking contributions 
from 
renormalization group running 
and anomaly mediation are canceled each other 
at the mirage messenger scale.
Then, 
pure modulus mediation appears, 
and hence the 
scale is regarded as the effective messenger scale of the soft masses. Thus, in the 
TeV scale mirage mediation scenario, the effective low messenger scale appears and 
the mass spectrum is controlled by modular weights and discrete parameters in the 
UV theory. 

The earlier study for the TeV scale mirage mediation in the 
minimal supersymmetric standard model (MSSM) has shown
the natural 
EWSB scenario \cite{Choi:2005hd, Kitano:2005wc, Choi:2006xb}. In the MSSM, relatively 
large stop mass and large $\tan\beta$ is favored in order to evade the lower bound 
of the lightest Higgs boson mass. On the other hand, the $\mu$ term is small in 
this scenario: It should satisfy the EWSB condition,
$\mu^2 \sim - m_{H_u}^2 - m^2_Z/2$, and the absolute value of the negative up-type 
Higgs soft mass squared, $m_{H_u}^2$, is not large due to the short renormalization 
group running from the TeV mirage scale. Thus, there are large mass difference 
between the Higgsino and other SUSY particles in the MSSM
case.

In the 
next-to minimal supersymmetric standard model (NMSSM), 
the situation is different; stop masses and $\tan\beta$ could be 
small if there is the sizable contribution to the lightest Higgs boson mass at 
the tree level additionally. Then, 
Higgsino 
mass is determined not only by 
$m_{H_u}$ but also by the down-type Higgs soft mass, $m_{H_d}$. Thus, the NMSSM 
can realize 
more diverse SUSY mass spectra than those in the MSSM. 
In this paper, we 
study the possible mass spectrum in the NMSSM with TeV mirage mediation 
and show the natural supersymmetric spectrum. This scenario is very interesting 
in the case of the lightest Higgs mass $\simeq 125$ GeV 
\cite{Chatrchyan:2012tx,HiggsATLAS}, especially. 

This paper is organized as follows. In the next section, we introduce the 
Lagrangian of the NMSSM and soft masses in the mirage mediation of the SUSY 
breaking. In Section 3, we discuss the EWSB in the NMSSM with TeV mirage 
mediation and show the samples of mass spectra. Section 4 is devoted to summary 
and discussion.

\section{Mirage mediation in the NMSSM}
\label{sec:model}

In this section, we briefly introduce the  SUSY breaking soft masses 
generated by the mirage mediation in the NMSSM. 
(For details of NMSSM, see reviews, e.g. \cite{Ellwanger:2009dp}.)

\subsection{NMSSM Lagrangian}

In the NMSSM, the singlet field $S$ is added to the MSSM. We consider the 
superpotential in the global SUSY for the singlet field, 
\begin{eqnarray}
  W &\supset& 
  \lambda S H_u H_d + \xi_F S + \frac{\mu_S}{2}S^2 +
\frac{1}{3} \kappa S^3,
  \label{eq:W_NMSSM}
\end{eqnarray}
and the relevant part of the soft-SUSY breaking term is given by
\begin{eqnarray}
  {\cal L}_{\rm soft} &\supset& 
  - m_{H_d}^2 |H_d|^2 - m_{H_u}^2 |H_u|^2 - m_{S}^2 |S|^2  \nonumber \\
&&   
- \left( \lambda A_\lambda S H_u H_d + \xi_F C_S S + \frac{1}{2}\mu_S B_S S^2
    + \frac{1}{3} \kappa A_\kappa S^3 + bH_u H_d + \rm{h.c.}\right).
\end{eqnarray}
The SUSY invariant Higgs mass, $\mu$, will be absorbed into the $S$ by its
shift without a loss of generality.

In this model, the parameters of the Higgs sector are given by 
($\lambda$, $\kappa$, $A_\lambda$, $A_\kappa$, $m_{H_d}^2$, $m_{H_u}^2$, 
$m_{S}^2$, $\xi_F$, $\mu_S$, $C_S$, $B_S$, $b$), and these are rewritten by 
following quantities:
\begin{eqnarray*}
  \lambda, \quad \kappa, \quad A_\lambda, \quad A_\kappa, 
           \quad m_{H_d}^2, \quad m_{H_u}^2, \quad m_{S}^2, \quad 
  \xi_F , \quad C_S , \quad b , \quad m_Z, \quad \tan\beta.
\end{eqnarray*}
The $\mu_{\rm eff}$ is defined by $\mu_{\rm eff} = \lambda v_s$ and it plays 
the role of $\mu$-parameter in the MSSM. Here, 
$\tan\beta =\langle H_u\rangle /  \langle H_d\rangle$ and
$v_s = \langle S\rangle$ in which $\langle\cdots\rangle$ denotes 
the
vacuum expectation value (vev). These vevs should satisfy the 
extremum conditions $\partial_i V = 0~(i=H_u,~H_d,~S)$; 
\begin{eqnarray}
\frac{\sin2\beta}{2} & = &
\frac{\mu_{\rm eff} (A_{\lambda} + \kappa v_s) + b + \mu_{\rm eff} \mu_S + \lambda \xi_F}
{m_{H_u}^2 + m_{H_d}^2 + 2 \mu_{\rm eff}^2 + \lambda^2 v^2},  \label{EW_condi_other1}  \\
 \frac{m_Z^2}{2} &=& - \mu_{\rm eff}^2 
    + \frac{m_{H_d}^2 - m_{H_u}^2 \tan^2\beta}{\tan^2\beta - 1},   \\
- m_S^2
&=& 
\nonumber
\lambda v^2 \left( \lambda  - \kappa \sin2\beta \right)
+ \kappa v_s  \left( 2\kappa v_s  + A_{\kappa } + 3 \mu_S \right) 
\\
&&
+ B_S \mu_S + \mu_S^2  + 2 \kappa  \xi _F 
- \lambda v^2 \sin2\beta \frac{\left( A_{\lambda } +  \mu_S \right)}{2v_s} +  \xi_F \frac{C_S +  \mu_S }{v_s} .
\label{EW_condi_other2}
\end{eqnarray}

\subsection{Soft masses in the mirage mediation}
Next, we introduce the SUSY breaking soft masses generated through mirage 
mediation. The model is given by 
\begin{eqnarray}
K &=& K_{\rm moduli}(\Phi^I + \bar{\Phi}^{\bar{I}}) + Z_i(\Phi^I + \bar{\Phi}^{\bar{I}})|\psi_i|^2 , \\
W &=& W_{\rm moduli}(\Phi^I) + W_{\rm NMSSM}(\psi) ,\\
f_a & = & {\cal F}_a(\Phi^I) = \sum_I d_I^a \Phi^I + l_a, 
\end{eqnarray}
where $\Phi^I$, $\psi_i$ and $l_a$ are moduli having SUSY breaking $F$-term, 
chiral matter fields and constant from heavy moduli, respectively. We shall 
consider the gauge coupling unification(GUT) at the GUT scale $M_X$, which is 
the cut-off scale we supposed\footnote{
One must include the Konishi and K\"ahler-Weyl anomalies in order to obtain 
the correct cut-off scale \cite{Dixon:1990pc}. 
}: 
$\langle {\cal F}_a (\Phi^I) \rangle \equiv \langle {\cal F} (\Phi^I)\rangle  = g_{\rm GUT}^{-2}$. We have taken
$M_{\rm Pl} = 2.4 \times 10^{18}$ GeV $\equiv 1$ and will use such a unit in 
this paper. The relevant scalar potential is given by
\begin{eqnarray}
V_F = e^K
\bigg[
\sum_{I=\Phi,~\psi}(D_I W) (\overline{D_J W})K^{I\bar{J}} -3|W|^2
\bigg].
\end{eqnarray}
Here $D_{I} W = (\partial_I K)W + \partial_I W$ and $K^{I\bar{J}}$ is the 
inverse of the K\"ahler metric
$K_{I\bar{J}}=\partial_I \overline{\partial_J} K$.

At just below $M_X$, $M_X^{-}$, the soft parameters are described by 
\cite{Choi:2004sx, Choi:2005ge}
\begin{eqnarray}
\nonumber
M_a (M_X^{-})& \simeq & M_0 + \frac{b_a}{16\pi^2}g^2_{\rm GUT} m_{3/2} , \\
\nonumber
A_{ijk} (M_X^{-}) & \simeq & 
\tilde{A}_{ijk} - \frac{1}{2} ( \gamma_i (M_X) + \gamma_j (M_X) + \gamma_k (M_X) )m_{3/2} , \\
m_i^2 (M_X^{-}) & \simeq &
\nonumber
\tilde{m}_i^2 -\frac{\dot{\gamma}_i (M_X)}{4}m_{3/2}^2 \\
\nonumber
&& 
+ 
\frac{1}{16\pi^2}m_{3/2}
\left(
\sum_{i=i,j,k} \sum_{j,k}|y_{ijk}|^2(M_X) \tilde{A}_{ijk} - 4 \sum_a \sum_{i=i,j,k} C_2^a(i) g_a^2 (M_X) M_0
\right),
\label{eq:softmass_Mx}
\\
\end{eqnarray}
including the anomaly mediation contribution. In Eq.(\ref{eq:softmass_Mx}),
\begin{eqnarray}
M_{0} &\equiv &
\frac{F^I \partial_I {\cal F}}{{\cal F} + \bar{\cal F}}, 
\label{mirageM}
\\
\tilde{A}_{ijk} &\equiv &
- F^I \partial_I \log \left( 
\frac{e^K Y_{ijk}}{Z_i Z_j Z_k}
\right), 
\label{mirageA}
\\
\tilde{m}_i^2 & \equiv & - F^I \bar{F}^{\bar{J}}\partial_I \bar{\partial}_{\bar{J}}\log(e^{-K_{\rm moduli}/3}Z_i),
\label{miragem}
\end{eqnarray}
are the gaugino mass, $A$-term and soft mass mediated by moduli at $M_X$, 
respectively. And $\gamma_i$ and $b_a$ are anomalous dimensions and 
coefficients of one-loop beta function of the gauge coupling $g_a$;
\begin{eqnarray}
\gamma_i (Q)
&=& \frac{d\log Z_i (Q)}{d\log Q} 
 =  \frac{1}{16\pi^2} \left(
4 \sum_a C_2^a(i) g_a^2(Q) - \sum_{j,k} |y_{ijk}|^2(Q)
\right), \\ 
b_a &=& -3 {\rm Tr}(T_a^2({\rm Adj})) + \sum_i {\rm Tr}(T_a^2(\psi_i)) 
= \left(-3, 1, \frac{33}{5} \right) \quad {\rm for~(N)MSSM}.
\end{eqnarray}
At the renormalization scale $Q$, gaugino masses, $A$-terms  
and scalar masses are given by \cite{Choi:2005uz, Endo:2005uy}\footnote{
We used the one-loop approximation of beta functions and anomalous
dimensions.}
\begin{eqnarray}
M_a & \simeq & M_0 \left( 1 - \frac{b_a}{8\pi^2}g_{a}^2(Q) \log \left(\frac{\Lambda}{Q}\right) \right), \\
A_{ijk} 
& \simeq & M_0 \left( a_{ijk} + (\gamma_i (Q)+ \gamma_j (Q)+ \gamma_k (Q) ) 
\log \left( \frac{\Lambda}{Q} \right) \right) , \\
m_i^2 & \simeq &
\nonumber
|M_0|^2
\bigg[
c_i
-\frac{g_Y^2 (Q)}{8\pi^2}Y_i (\sum_k c_k Y_k) \log\left(
\frac{M_X}{Q}
\right)
 \\
&&
+ \left( 2 \gamma_i (Q) - \dot{\gamma}_i (Q)
\log\left(
\frac{\Lambda}{Q}
\right)
\right)
\log\left(
\frac{\Lambda}{Q}
\right)
\bigg ], 
\end{eqnarray}
where 
\begin{eqnarray}
a_{ijk} &=& \frac{\tilde{A}_{ijk}}{M_0}, \qquad
c_i = \frac{\tilde{m}_i^2}{|M_0|^2}.
\end{eqnarray}
For
large Yukawa couplings $y_{ijk}$, it is required that
\begin{eqnarray}
a_{ijk} &=& 1 ~~~{\rm and}~~~ c_i + c_j  + c_k = 1 , 
\label{eq:mirage_condi}
\end{eqnarray}
in order to obtain the above result. 
In other words, this condition implies that they have the same scaling on 
the moduli $\Phi^I$:
\begin{eqnarray}
{\rm Re}({\cal F}) \sim
e^{-K_{\rm moduli}}Z_{H_u}Z_{Q3}Z_{\bar{U}3} \sim
e^{-K_{\rm moduli}}Z_S Z_{H_u}Z_{H_d} 
\label{Ycond1}
\end{eqnarray}
for e.g. large $y_t$ and $\lambda$.

The mirage messenger scale $\Lambda$ is given by
\begin{eqnarray}
\Lambda \equiv M_X \left(\frac{m_{3/2}}{M_{\rm Pl}}\right)^{\alpha /2} , 
\end{eqnarray}
using
\begin{eqnarray}
\alpha \equiv \frac{m_{3/2}}{\log(M_{\rm Pl}/m_{3/2})M_0}, 
\end{eqnarray}
which is the ratio of anomaly mediation to 
moduli mediation\footnote{
In general, one should replace $m_{3/2}$ with $F^{\phi}$, where 
$F^{\phi} = m_{3/2} + \partial_I K F^I/3$. If there were the no-scale moduli 
breaking SUSY, the anomaly mediation would be suppressed.
}.
As shown above, pure modulus mediation appears at $\Lambda$, not at the Planck 
scale, and hence this mediation is called the mirage mediation. 
The
$\Lambda$ 
behaves the effective cut-off scale of modulus-mediated soft masses.

\subsection{Model details}

For a concreteness, we shall consider the KKLT-like model 
\cite{Kachru:2003aw, Abe:2005rx, Choi:2006xb}, with a single light volume modulus 
$T$\footnote{
We have taken the normalization $4\pi T = V_{\rm CY}^{2/3}$, where $V_{\rm CY}$ is a six 
dimensional compactification volume of the Calabi-Yau space in the Einstein frame.
} 
and a stabilized heavy modulus ${\cal S}$ whose vev is denoted as 
${\cal S}_0 = (4\pi g_s)^{-1}$, where $g_s$ is the string coupling:
\begin{eqnarray}
K &=& -3\log(T+\bar{T}) + Z_i |\psi_i|^2 , \\
W &=& 
A_0 e^{-8\pi^2 l_0 {\cal S}_0} -Ae^{-8\pi^2 (k_h T + l_h {\cal S}_0)} + W_{\rm NMSSM}
\label{eq:WNMSSM}
,\\
f_{v} &=& k_v T + l_v {\cal S}_0 ,
\end{eqnarray}
where $f_v$ is the unified gauge coupling function in the visible sector.
In terms of string theory, one would find, for instance, 
\begin{eqnarray}
k_h = \frac{w_h}{N}, \qquad k_v = w_v.
\end{eqnarray}
Here $N$ can originate from the $SU(N)$ gaugino condensation \cite{Dine:1985rz} 
or instanton for $N=1$ \cite{Blumenhagen:2009qh}. 
The 
$w_v \in {\mathbf Z}$ and 
$w_h \in {\mathbf Z}$ would denote the wrapping number of a visible/hidden 
D-brane on the relevant cycle, which experiences such non-perturbative effects. 
The parameters $l_i \in {\mathbf Z}$ in the visible/hidden gauge coupling 
function implies, e.g., the flux $F_2$ and geometric curvature ${\cal R}_2$:
$4\pi^2 l_i = \int_{C_i^{(4)}} {\rm Tr}(F_2 \wedge F_2) + {\rm Tr}({\cal R}_2 \wedge {\cal R}_2)$ on the D7-brane wrapping on the four cycle $C_i^{(4)}$. Hence $l_h$ would be 
$l_h'/N$, where $l_h' \in {\mathbf Z}$ is given by the such flux and geometric curvature.
The first term in the superpotential may come from heavy moduli stabilized by 
flux. 
For later convenience we parametrized as
$\langle W_{\rm flux} \rangle = \langle \int G_3 \wedge \Omega \rangle \equiv A_0 e^{-8\pi^2 l_0 {\cal S}_0} \sim 10^{-13}$ .

In order to break the SUSY and to realize de Sitter/Minkowski vacuum, 
a sequestered SUSY-breaking anti-brane on a top of the warped throat is viable\footnote{
A dynamical SUSY breaking is also possible \cite{Lebedev:2006qq, Dudas:2006gr, Abe:2006xp, Kallosh:2006dv}.
}; the total scalar potential is given by\footnote{Thus the modulus $T$ is 
stabilized near SUSY location 
$D_T W \simeq 0$ and $\langle V_F \rangle \simeq -3 m_{3/2}^2 \sim -3|W_0|^2$, 
where $W_0 = A_0 e^{-8\pi^2 l_0 {\cal S}_0}$. See also \cite{Higaki:2011me}.
}
\begin{eqnarray} 
V = V_F + V_{\rm lift}  ,
\end{eqnarray} 
where
\begin{eqnarray} 
V_{\rm lift} = \epsilon (T+\bar{T})^{-2} .
\end{eqnarray}
As a result, the cosmological constant is almost vanishing via the fine tuning 
\begin{eqnarray}
&& \epsilon \sim   A_0^2 e^{-16\pi^2 l_0 {\cal S}_0} , \\
&& {\rm where}~~~
\epsilon^{1/2} \propto \exp \left[- \frac{2h}{3f} 8\pi^2 {\cal S}_0
\right] \equiv e^{-8\pi^2 l_0 {\cal S}_0}.
\end{eqnarray}
Here $\epsilon^{1/4}$ is the minimum of warp factor on the Klebanov-Strassler 
throat \cite{Klebanov:2000hb} on which the anti-brane is sitting,
$h = - \int_{C_3} H_3^{\rm NSNS} \in {\mathbf N}$ and $f = \int_{C_3} F_3^{\rm RR} \in {\mathbf N}$ \cite{Giddings:2001yu}.

One finds the followings in the vacuum with $m_{3/2} =O(10)$ TeV:
\begin{eqnarray}
k_h  T + l_h {\cal S}_0  &\simeq & l_0 {\cal S}_0 \simeq \frac{1}{8\pi^2} \log (M_{\rm Pl}/m_{3/2}) \simeq \frac{1}{2} ,\\ 
\frac{F^T}{T+\bar{T}} & = &  2 \frac{m_{3/2}}{8\pi^2 k_h(T+\bar{T})} 
\simeq \frac{m_{3/2}}{\log (M_{\rm Pl}/m_{3/2})}\frac{l_0}{l_0-l_h} ,
\\
M_0  & = & k_v \frac{F^T}{T + \bar{T}} 
\frac{T+\bar{T}}{k_v (T+\bar{T}) + l_v (S+\bar{S})}  \nonumber \\
&=&   \frac{F^T}{T + \bar{T}}\frac{k_v(l_0 - l_h)}{k_v (l_0 - l_h) + l_v k_h}  \nonumber \\
& = & \frac{m_{3/2}}{\log(M_{\rm Pl}/m_{3/2})} \frac{k_v l_0}{k_v (l_0 - l_h) +l_v k_h} ,
\end{eqnarray}
where $m_{3/2} \simeq e^{K/2}A_0 e^{-8\pi^2 l_0 {\cal S}_0} \sim A_0 e^{-8\pi^2 l_0 {\cal S}_0}$. On the other hand, the visible gauge coupling function indicates 
\begin{eqnarray}
{\rm Re}(f_v) = k_v \langle T \rangle + l_v {\cal S}_0 = g^{-2}_{\rm GUT} \simeq 2,
\end{eqnarray}
in the vacuum, hence it will imply
\begin{eqnarray}
&&  
\frac{ k_v \left(l_0-l_h\right)+k_h l_v}{k_h l_0} \simeq
\frac{8\pi^2}{g_{\rm GUT}^2 \log({M_{\rm Pl}}/{m_{3/2}})} = 4 -5.
\end{eqnarray}
Then the ratio $\alpha$ becomes
\begin{eqnarray}
\alpha \equiv \frac{m_{3/2}}{\log(M_{\rm Pl}/m_{3/2})M_0} \simeq
\frac{k_v (l_0 - l_h) +l_v k_h}{k_v l_0} 
&\simeq & \frac{k_h}{k_v} \times ( 4 -5 ) .
\label{eq:alpha_amb}
\end{eqnarray}
Now, $k_h/k_v = w_h/(w_v N)$ will be a rational number at any rate.
In the global IIB orientifold models, the cut-off scale of 4D theory will be 
given by the compactification scale in the Einstein frame:
\begin{eqnarray}
M_X \equiv M_{\rm KK} \sim \frac{M_{\rm Pl}}{8\pi {\cal S}_0^{1/4} \langle T \rangle} . 
\end{eqnarray}
When ${\cal S}_0$ and $\langle T\rangle$ are of $O(1)-O(10)$, one would find 
$M_X \sim 10^{16}$ GeV. If, for example, $w_v=1,~w_h=2$ and $N =5$, one obtains 
\begin{eqnarray}
\alpha \simeq 2, \qquad \Lambda = O(1) {\rm TeV},
\end{eqnarray}
for $M_X = O(10^{16})$ GeV and $m_{3/2} =O(10)$ TeV.

Hereafter we will consider the following as the superpotential in Eq.(\ref{eq:WNMSSM}),
\begin{eqnarray}
\nonumber
W_{\rm NMSSM} 
&=& \lambda S H_u H_d 
+ B_1 \exp[-8\pi^2 (k_h T + (l_0 + l_h){\cal S}_0)] S \\
\nonumber
&& + B_2 \exp[-4\pi^2 (k_h T + (l_0 + l_h){\cal S}_0)] S^2 \\
&& + B_3 \exp[-8\pi^2 (k_{\kappa} T + l_{\kappa} {\cal S}_0)] S^3 ,
\label{WNMSSMmirage}
\end{eqnarray}
assuming that $S$ can have a charge of a Peccei-Quinn (PQ)
$U(1)_{\rm PQ}$ symmetry which is broken down at the non-perturbative level 
or that $S$ is moduli which possesses an approximate shift symmetry. 
Hence polynomial in $S$ would appear only at the non-perturbative level; 
all the terms will be naturally much smaller than of $O(1)$ in the Planck 
unit, i.e. $\kappa$ will be negligible. In this setup, one will obtain
\begin{eqnarray}
\mu_S \sim \sqrt{| \xi_F |} \sim B_S \sim C_S \sim M_0
\end{eqnarray}
by choosing the above exponent (see appendix \ref{non-pert_kappa}).
If there were non-perturbative $\mu$-term similarly, the magnitude will also be 
supposed to be of $O(M_0)$\footnote{
One can construct such a model with Affleck-Dine-Seiberg (ADS) superpotential,  
taking ${\cal O}= H_u H_d$ in Appendix \ref{ADS} \cite{Choi:2006xb}.}.

\subsubsection{Matter K\"ahler potential}

With world volume fluxes to realize chiral matter on the branes, it is expected 
that the K\"ahler potential is corrected by the flux like a case of the gauge 
coupling such that $Z(T+\bar{T}) \to Z(T+\bar{T} + l ({\cal S}_0+\bar{{\cal S}}_0) )$.
However, the correction highly depends on the model and computation is difficult. 
Hence there exists just the scaling argument on K\"ahler (volume) moduli in the 
Calabi-Yau compactifications \cite{Conlon:2006tj}. Only for simple cases of 
toroidal compactifications, it is possible to discuss the exact flux contributions 
to the K\"ahler potential at the present status \cite{Cremades:2004wa}.
Thus we will consider the phenomenological K\"ahler potential in this paper:
\begin{eqnarray}
e^{-K_{\rm moduli}/3}Z_i 
= (T+\bar{T} + l_i ({\cal S}_0+\bar{{\cal S}}_0)/k_i)^{r_i} .
\label{matterKahler}
\end{eqnarray}
Then one finds
\begin{eqnarray}
\tilde{A}_{ijk} &=& 
(s_i + s_j + s_k)M_0, \qquad
s_i =
\frac{k_i \left(k_v (l_0-l_h ) + k_h l_v \right)} 
{k_v\left(k_i (l_0-l_h )+k_h l_i\right)}r_i
, \\
\tilde{m}_i^2 
 &=&  c_i |M_0|^2 ,  \qquad \qquad \qquad
c_i =
\left(\frac{k_i \left(k_v (l_0-l_h ) + k_h l_v \right)} 
{k_v\left(k_i (l_0-l_h )+k_h l_i\right)}\right)^2
r_i  .
\end{eqnarray}
Thus, for the mirage mediation in the large Yukawa sector, the condition 
\begin{eqnarray}
a_{ijk}=s_i + s_j + s_k = 1 ~~~{\rm and}~~~ c_i + c_j + c_k = 1, 
\end{eqnarray}
is required for the successful mirage mediation. Choosing all $k_i = k_v$ and 
$l_i = l_v$ as the gauge coupling in the visible sector, the case of
\begin{eqnarray}
s_i = c_i = r_i \, \leq 1
\end{eqnarray} 
is viable.
We will use this relation later, 
especially for the large Yukawa sector ($y_t$ and $\lambda$),
as in earlier studies of the mirage mediation,
while a general K\"ahler potential will be considered for the other sector.

\subsubsection{On modular weight $r_i$}
Here, we briefly discuss the value of modular weights. In IIB orientifold 
supergravity (SUGRA), it was discussed that there would be possibilities 
relevant to the K\"ahler modulus \cite{Conlon:2006tj}:
\begin{eqnarray}
r_i = 0, ~\frac{1}{6},~\frac{1}{3},~\frac{1}{2}, 1 .
\end{eqnarray}
For instance, chiral matter localized on the same stack of the magnetized 
D7-brane will have $r_i = 1/3$ while those of position moduli and Wilson 
line moduli on the D7-brane will correspond to $r_i = 1$ and $r_i =0$ respectively.
On the D3-brane which is sitting on a singularity (a fractional D3-branes), 
there would be matter fields charged under the gauge group if we have such branes. 
Including multiple D7 and D3-branes, $r_i = 0$ (D3-D3 state or a state on the 
vanishing cycle) or $r_i = 1/2$ (D3-D7 state or also D7$_i$-D7$_j$ state on the 
two-cycle intersection between the different stacks of D7-branes whose triple 
intersection is a single point) will be found. Furthermore, if any pair of three 
stacks of the D7-branes, which are wrapping on the different cycle, are 
intersecting at the same two-cycle, $r_i = 1/6$ would be realized for the 
relevant matter.
With multiple moduli, depending on the model, there could be other possibilities 
effectively\footnote{ 
In examples of toroidal compactifications with chiral matter localized between 
two D7-branes \cite{Cremades:2004wa}, there will be a model with 
$K= -\sum_{I=1}^3 \log (T_I +\bar{T}_I) + (T_1+\bar{T}_1)^{-1/2}|\psi_i|^2$ up 
to world volume flux contribution. Thus one has 
$e^{-K/3}Z_i = (T_1+\bar{T}_1)^{-1/6}\prod_{I=2}^3 (T_I +\bar{T}_I)^{1/3}$. 
The result of SUSY breaking will depend on the 
model, 
however, such a background can be symmetric under moduli and hence it could be expected that
$F^{T_1}/(T_1+\bar{T}_1) = F^{T_2}/(T_2+\bar{T}_2) = F^{T_3}/(T_3 + \bar{T}_3) \equiv M_0$ 
and then $m_i^2 = |M_0|^2/2$ from moduli. This means that effectively $r_i = 1/2$ is obtained for the relevant soft scalar mass. However if one obtains 
different 
$F$-terms from the above one, it could be found that $r_i \neq 1/2$ effectively.
}.

In heterotic string model, one may find at the leading of string coupling\footnote{
In heterotic-M theory, there will be a correction by dilaton
$Z = [1/(T+\bar{T}) + \beta/({\cal S}_H+\bar{{\cal S}}_H) ]$.
}
\begin{eqnarray}
r_i({\cal S}_H) = \frac{1}{3}
\end{eqnarray}
on the string dilaton ${\cal S}_H$, which would be light and have the K\"ahler 
potential $K=-\log({\cal S}_H+\bar{{\cal S}}_H)$. For fields propagating bulk 
(untwisted sector on the orbifold), they will have
\begin{eqnarray}
r_i(T_H) = 0
\end{eqnarray}
on the K\"ahler modulus with the K\"ahler potential $K=-3\log(T_H+\bar{T}_H)$.
Based on the string duality between heterotic string and IIB string, 
the result about modular weight on $T$, which is relevant to SUSY breaking scalar 
mass, would be obtained\footnote{
In a twisted sector on the heterotic $Z_{N}$ orbifold,
a state there will have $Z_i \sim \prod_I(T_I +\bar{T}_I)^{-1 + k_I^i/N}$, where
$k_I^i$ is a positive integer. The K\"ahler potential $K= -\log(V_{\rm CY})$ is 
somewhat complicated; the result depends on the model, and unfortunately the 
viable moduli stabilization mechanism is less known than that of type IIB case.
}.

At any rate, we expect that the perturbative physical Yukawa coupling,  
\begin{eqnarray}
|y_{ijk}|^2  \sim \frac{e^{K_{\rm moduli}}}{Z_i Z_j Z_k} \sim (T+\bar{T})^{-(r_i + r_j + r_k)},
\end{eqnarray}
is at most completely localized (constant coupling) or spreading as the gauge 
coupling on the relevant geometry (volume-suppressed coupling),and hence an 
effective modular weight $r_i$ could lie on the range with 
\begin{eqnarray}
0 \leq r_i \leq 1,
\end{eqnarray}
taking various values. A case with $r_i < 0$ will be an implausible model in 
the extra dimension because a Yukawa coupling is growing as the $T$ (volume) 
becomes large. Therefore we will not consider such cases.

For phenomenological reasons, the possibility of $r_i \neq c_i$ would be 
interesting. For example, the large $c_{q_{1,2}}$ achieves the hierarchical 
squark mass matrix avoiding FCNC. In that case, the constraints are relaxed 
because there are no gluino-squark production at LHC. Moreover, small $c_{l_i}$ 
realize light slepton which may be favored by the muon $g-2$ anomaly
\cite{muong2Hagiwara,muong2Davier}. Here, we 
demonstrate to construct the heavy squark mass spectrum, considering following 
cases in general case with the K\"ahler potential Eq.(\ref{matterKahler}): 
\begin{eqnarray}
\nonumber
c_{Q_{1,2}} &=& c_{\bar{U}_{1,2}} = c_{\bar{D}_{1,2}} = c_{L_{1,2,3}} = c_{\bar{E}_{1,2,3}} \equiv c_M > 1, 
\\
c_{Q_3} &=& c_{\bar{U}_3} = c_{\bar{D}_3} = \frac{1}{2},~
c_{H_u} = 0,~c_{H_d} = 1,~c_{S}= 0, \\
&& s_{H_u} + s_{Q_3} + s_{\bar{U}_3} = s_{S} +s_{H_u} + s_{H_d} = 1, \qquad
a_M \geq 1,
\end{eqnarray}
with $r_M < 1$. 
Here $M$ denotes the matter except for third quark multiplets 
$Q_3,~\bar{U}_3,~\bar{D}_3$ and the case of $c_i = s_i = r_i$ for third 
generation quarks and Higgs sector is used.
This is because the above condition is not required for the small Yukawa coupling sector,
where one can obtain
\begin{eqnarray}
r_M &=& \frac{1}{4c_{M}} < 1, \qquad
\frac{k_{M} \left(k_v (l_0-l_h ) + k_h l_v \right)} 
{k_v\left(k_{M} (l_0-l_h )+k_h l_{M}\right)} = 2c_{M} \\ 
&& 
\to 
a_{H_u ij} = 1, \qquad a_{H_d i j} = 2 , \qquad
a_{H_u Q_3 \bar{U}_3} = a_{S H_u H_d} = 1 
.
\end{eqnarray}
The parameters in the relevant sector
\begin{eqnarray}
{\cal S}_0 &=& 1,~k_h = \frac{1}{2},
~l_0 = \frac{1}{2},~l_h=-2, \\
k_v &= & 1,~ l_v = -3
,~l_{M}=-39,~k_{M}=8 ,
\end{eqnarray}
lead to\footnote{
For simplicity, we will not consider a GUT relation.
}
\begin{eqnarray}
c_{M }= 8,  
\end{eqnarray}
$Z_i > 0,~ \langle T \rangle \simeq 5 , ~ g_{\rm GUT}^{-2} \simeq 2,~
{\rm and}~ M_X \sim 1.9 \times 10^{16} {\rm GeV}$; the larger $T$ becomes, 
the smaller $M_X$ is. For another example, if we replace $l_{M}=-39$ with 
$l_{M}=-38$, $c_{M }= 4$. These are just samples and the $c$, of course, 
can take various other values. 

\section{TeV scale mirage mediation in the NMSSM}
\label{sec: TeV mirage in NMSSM}

Hereafter we shall consider the TeV scale mirage mediation ($\alpha \sim 2$) 
to focus on the natural EWSB. The relevant soft parameters in the Higgs sector 
for TeV scale mirage mediation are exhibited in Appendix \ref{softmass_TeVmirage}.

At first, we see that the TeV mirage mediation scenario works to achieve the 
natural EWSB in the MSSM \cite{Choi:2005hd, Kitano:2005wc, Choi:2006xb}.
In the (N)MSSM, the electroweak scale are determined by the equation
\begin{eqnarray}
 \frac{m_Z^2}{2} &=& - \mu_{\rm (eff)}^2 
    + \frac{m_{H_d}^2 - m_{H_u}^2 \tan^2\beta}{\tan^2\beta - 1}  \nonumber \\
&\sim& - \mu_{\rm (eff)}^2 
    - m_{H_u}^2 + \frac{m_{H_d}^2}{\tan^2\beta},
\label{eq:ew_condition}
\end{eqnarray}
where the last relation holds for a moderately large $\tan\beta$. The 
$m_{H_u}^2$ would receive a correction through the renormalization group 
running, $\sim -3y_t^2/(8 \pi^2) (m_{Q3}^2 + m_{U3}^2 + |A_t|^2) 
\log(\Lambda_{\rm mess}/M_{\tilde{t}})$, where $\Lambda_{\rm mess}$ and $M_{\tilde{t}}$ 
are the messenger scale and the stop mass scale, respectively. Since the 
correction is proportional to a logarithmic factor, 
$\log(\Lambda_{\rm mess}/M_{\tilde{t}})$, the low messenger scale suppresses 
this contribution and helps to realize the correct EWSB without serious fine 
tuning.

In the MSSM, the large $\tan\beta$ is favored for the lightest Higgs boson mass
to be large at the tree level. Moreover, the large radiative correction from 
relatively heavy stops are also needed to evade the Higgs mass bound by LEP 
experiments. In high scale messenger models, however, the heavy stops develops a 
large logarithmic correction to $m_{H_u}^2$ in Eq.(\ref{eq:ew_condition}) and 
therefore it causes the fine tuning problem. That is called the SUSY little 
hierarchy problem.
In the TeV scale mirage mediation, the running correction from the mirage 
messenger scale is not so large because of the effective low messenger scale. 
Thus, the stop can be relatively heavy without fine tuning and lift the 
lightest Higgs boson mass up.

Actually, proper assignment of discrete 
parameters of $c_{H_u}$ and $c_{Q_3,\bar{U}_3}$ leads to a natural solution of 
Eq.(\ref{eq:ew_condition}) and satisfy the Higgs boson mass mass constraint 
by LEP:
\begin{eqnarray}
 c_{H_u} = 0 ~~~{\rm and}~~~ c_{Q_3} + c_{\bar{U}_3} = 1, 
\end{eqnarray}
where the soft masses are obtained as
\begin{eqnarray}
 m_{H_u} &=& \sqrt{c_{H_u}}M_0 + \mathcal{O}(M_0/\sqrt{8 \pi^2})
          = \mathcal{O}(M_0/\sqrt{8 \pi^2})  ,  \\ \nonumber
 m_{\tilde{t}_{L,R}} 
         &=& \sqrt{c_{t_{L,R}}}M_0 + \mathcal{O}(M_0/\sqrt{8 \pi^2})
          = \mathcal{O}(M_0),
\end{eqnarray}
at the mirage messenger scale. The $\mathcal{O}(M_0/\sqrt{8 \pi^2})$ in above 
equations denotes higher order effects. From Eq.(\ref{eq:ew_condition}), 
$\mu$ term is also of $\mathcal{O}(M_0/\sqrt{8 \pi^2})$.

As described above, the TeV scale mirage mediation in the MSSM achieves the 
correct EWSB naturally. In the mass spectrum, there exists the large mass gap 
between Higgsino masses being of $\mathcal{O}(M_0/\sqrt{8 \pi^2})$, and other 
SUSY particle masses being of $\mathcal{O}(M_0)$, in order to solve the SUSY 
little hierarchy problem.

On the other hand, there are no SUSY little hierarchy problem in the NMSSM 
if there is a large additional tree level contribution to the lightest 
Higgs boson mass. The situation would be realized for small $\tan\beta$. In 
that case, heavy stop masses is not 
required for the Higgs mass bound. Furthermore, although the $m_{H_u}$ in 
Eq.(\ref{eq:ew_condition}) is small due to a little deviation from the pure 
modulus mediation, the value of $m_{H_d}$ contribution can be from of 
$\mathcal{O}(M_0/\sqrt{8 \pi^2})$ to of $\mathcal{O}(M_0)$, depending on the 
parameter $c_{H_d}$. Thus, the $\mu$ term will lie on the range from of 
$\mathcal{O}(M_0/\sqrt{8 \pi^2})$ to of $\mathcal{O}(M_0)$. In particular, 
$\mu$ term will be the same order of other SUSY particle's masses in the case 
of $c_{H_d} = 1$, and the mass spectrum will be compressed.

\subsection{Possible mass spectrum}
Here, we briefly discuss the possible 
mass
spectra in the NMSSM with TeV scale mirage 
mediation scenario. If we consider the successful TeV mirage mediation scenario 
in which the soft mass are described by the boundary conditions in 
Eq.(\ref{eq:mirage_condi}), the following condition is satisfied, 
\begin{eqnarray}
c_{H_u} + c_{Q_3} + c_{\bar{U}_3} = 1,
\end{eqnarray}
because the top Yukawa coupling is large. And we also impose 
\begin{eqnarray}
c_{H_u} + c_{H_d} + c_{S} = 1,
\end{eqnarray}
because $\lambda$ should be also large in order for a large additional 
contribution to lightest Higgs boson mass. For simplicity, we will consider 
the small $\kappa$ region (and its string theoretical motivation is discussed 
in Appendix B). If one takes large $\kappa$, $c_{S} = 1/3$ is required for 
successful mirage mediation.

In the following part, we study the possible mass spectrum in the NMSSM with 
TeV mirage mediation, considering the degenerate spectrum and hierarchical 
spectrum as the benchmark points.


\begin{itemize}

\item {\bf Degenerate spectrum }

\end{itemize}

To achieve the EWSB and degenerate spectrum in TeV scale mirage mediation, the 
absolute value of $m_{H_u}^2$ at the mirage messenger scale should be small. 
Then, as the sample points, we take the following assignment of discrete 
parameters:
\begin{eqnarray}
 \left( c_{H_u} , c_{H_d}, c_S \right) &=& \left( 0, 1, 0 \right) 
                ~~~{\rm and}~~~ c_{Q_3} = c_{\bar{U}_3} = 1/2.
\end{eqnarray}
To focus on the degenerate SUSY spectrum, we consider only small $\tan\beta$ 
region in this section.

The gaugino would be heavier than stops and the mass difference between gluino 
and stop is greater than $(1-\sqrt{c_{Q_3,\bar{U}_3}})M_0$, which will become 
the minimum value at our sample point, following Eq.(\ref{eq:mirage_condi}). 
Then, the mass difference is greater than top mass if 
$m_{\tilde{g}} \sim M_0 \gtrsim 600$ GeV. The branching ratio of gluino decay 
depends also on the other squarks masses determined by 
$c_{Q_{1(2)}},c_{\bar{U}_{1(2)}}$ and $c_{\bar{D}_{1(2,3)}}$. In the case of 
$c_{Q_i,\bar{U_i},\bar{D_i}} = c_{L,\bar{E}} = c_{Q_3,\bar{U}_3}$, dominant 
decay branches are 
$\tilde{g} \to j \tilde{q}  \to jj \tilde{\chi}^0_1$ and $ jj \tilde{\chi}^\pm_1$. 
Eventually $\tilde{\chi}^\pm_1$ decays into $jj \tilde{\chi}^0_1$ or 
$l \nu_l \tilde{\chi}^0_1$.

\begin{figure}[htbp]
 \begin{center}
\vspace{-1.7cm}
  \includegraphics[width=100mm]{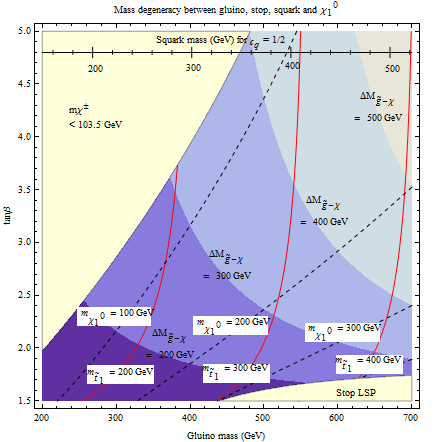} 
  \includegraphics[width=100mm]{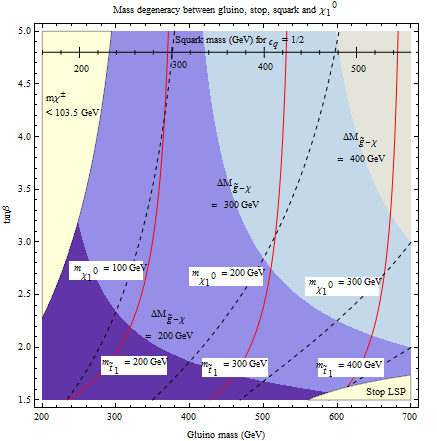}
\vspace{-1.0cm}
 \end{center}
 \caption{\small Contours of the mass difference between gluino and lightest neutralino 
$\Delta M_{\tilde{g}-\chi}$ for $\log(\Lambda/M_{\rm SUSY}) = 1$ and $4$.
The solid (dotted) line shows the lighter stop mass $m_{\tilde{t}_1}$ 
(the lightest neutralino mass $m_{\chi_1^0}$). The left upper region excluded by 
the chargino mass limit of $103.5$ GeV \cite{charginoLEP}. 
the $\mu_{\rm eff} < 100$ GeV and the region. The right bottom region indicates 
the Stop LSP region. In this figure the relevant parameter are chosen as
$\left( c_{H_u} , c_{H_d}, c_S \right) = \left( 0, 1, 0 \right)$, 
$c_{Q_3} = c_{\bar{U}_3} = c_{\bar{D_3}} = 1/2$ and $\lambda =0$.
}  \label{fig1}
\end{figure}

Furthermore, the difference between gluino mass and $\mu_{\rm eff}$ is 
determined by Eq.(\ref{eq:ew_condition}), which depends only on $M_0, \tan\beta$ 
and $c_{H_u, H_d}$. It is particularly worth noting that the degeneracy of the 
successful TeV scale mirage mediation in the NMSSM are mainly determined by this 
relation. This is because the mass difference between gluino and lightest 
neutralino is not smaller than the difference between gluino mass and 
$\mu_{\rm eff}$. In Fig.\ref{fig1}, we show the mass difference between gluino 
and lightest neutralino in the case of $(c_{H_u}, c_{H_d}, c_{S}) = (0, 1, 0)$ 
and $\lambda = 0$. In the $\lambda = 0$ case, singlino decouples from the 
neutralino mixing and the mass of MSSM neutralinos are determined only by the 
$M_0$ and $\tan\beta$. For $\lambda \ne 0$, the lightest neutralino mass will 
be smaller than that in the case of $\lambda = 0$ via the mixing with the 
singlino. Hence we take $\lambda = 0$ for showing the minimum mass difference 
in the scenario. 


In Fig.\ref{fig1}, the boundary of the shaded region shows the contour of 
constant gluino-neutralino mass difference denoted by $\Delta M_{\tilde{g}-\chi}$. 
The solid (dotted) line shows the lighter stop mass (the lightest neutralino mass). 
We have taken the mirage scale as $\log(\Lambda/M_{\rm SUSY}) = 1$ and $4$ 
(these correspond to $\Lambda = \mathcal{O}(1)$ TeV and $ \mathcal{O}(10)$ TeV) 
in Fig.\ref{fig1}. This is because the mirage scale has an ambiguity depending on 
the moduli stabilization. For $\log(\Lambda/M_{\rm SUSY}) = 4$, the running 
contribution to the $m_{H_u}$ is larger than that in the case of 
$\log(\Lambda/M_{\rm SUSY}) = 1$, and thus the $\mu_{\rm eff}$ becomes larger; the 
gluino-lightest neutralino mass difference consequently small. In the small 
$\tan\beta$ region, the mass difference is small because the $m_{H_d}$ contribution 
in Eq.(\ref{eq:ew_condition}) isn't suppressed by $\tan\beta$. 
However, in the small $\tan\beta$ region $(\tan\beta \lesssim 2)$, the top Yukawa 
coupling might become too large at the GUT scale. In Fig.\ref{fig1}, we take 
$c_{Q_i,\bar{U_i},\bar{D_i}} = c_{L,\bar{E}} = c_{Q_3,\bar{U}_3}$. But first and 
second squarks can be heavy without changing the figure drastically, taking $c_M > 1$.
Since we have ignored two-loop renormalization group effects, there will be, for 
instance, of $O(M_0^2/8\pi^2)$ uncertainty on soft scalar masses. Thus, 
a large $c_M$ increases such uncertainties.

\begin{itemize}

\item {\bf Hierarchical spectrum }

\end{itemize}

As the sample point for the hierarchical spectrum, we have taken the following 
assignment of discrete parameters:
\begin{eqnarray}
 \left( c_{H_u} , c_{H_d}, c_S \right) &=& \left( 0, 0, 1 \right) 
                ~~~{\rm and}~~~ c_{Q_3} = c_{\bar{U}_3} = 1/2.
\end{eqnarray}
In this case, $\mu_{\rm eff} = \mathcal{O}(M_0/\sqrt{8 \pi^2})$ while 
stop and gluino masses are the order of $M_0$ because $m_{H_d}^2$ is 
also the order of $M_0/\sqrt{8 \pi^2}$. Then, the theory can avoid 
serious tuning even if $M_0$ is around $1$ TeV and $\tan \beta$ is of 
${\cal O}(1)$\footnote{ 
Due to the smallness of the $\mu_{\rm eff}$, 
there would be a tuning to obtain the appropriate $\tan\beta$ in 
the RHS of Eq.(\ref{EW_condi_other1}). 
But it is around $10 \%$, then, would be not so serious. 
}.

\begin{figure}[htbp]
 \begin{center}
  \includegraphics[width=100mm]{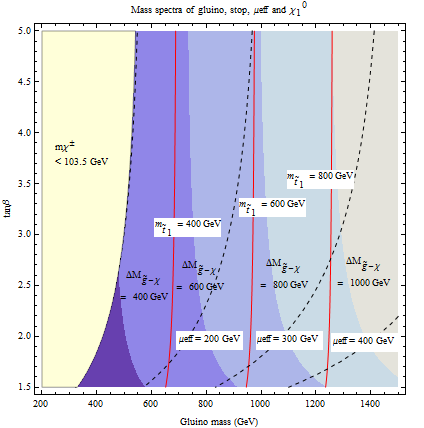} 
\vspace{-1.0cm}
 \end{center}
 \caption{\small Contours of the mass difference between gluino and lightest neutralino 
$\Delta M_{\tilde{g}-\chi}$ for $\log(\Lambda/M_{\rm SUSY}) = 1$.
The relevant parameters are given by
$\left( c_{H_u} , c_{H_d}, c_S \right) = \left( 0, 0, 1 \right)$,
$c_{Q_3} = c_{\bar{U}_3} = c_{\bar{D_3}} = 1/2$ and $\lambda =0$.
 }  \label{fig2}
\end{figure}

We also show the contour plot of mass spectra with 
$(c_{H_u}, c_{H_d}, c_{S}) = (0, 0, 1)$ in Fig.\ref{fig2}. 
Unlike in Fig.\ref{fig1}, the dotted line shows the contour of constant 
$\mu_{\rm eff}$ in Fig.\ref{fig2}. 
The right upper 
region in Fig.\ref{fig2} corresponds to the MSSM-like spectrum region. In the 
region, of course, there is also no serious fine tuning as well as the MSSM 
case. If $\tan\beta$ is not too large, the lightest Higgs boson mass can be 
heavier than the MSSM case because there is the additional contribution to 
the Higgs boson mass\footnote{ 
See also \cite{Kobayashi}.
}.
In the NMSSM, 
the mass difference between gluino and lightest 
neutralino can be greater than the result plotted in Fig.\ref{fig2} if the 
singlino is main component of the lightest neutralino.

In this subsection, we have discussed the features of mass spectra in the NMSSM 
with TeV scale mirage mediation, considering the degenerate spectrum and 
hierarchical spectrum. Although other values of discrete parameters can be taken, 
e.g. $(c_{H_u}, c_{H_d}, c_{S}) = (1/6, 5/6, 0)$, the features of such 
intermediate region of discrete parameter space can also be understood by above 
discussion. In the next subsection, we will discuss the experimental constraints 
for parameter space in these figures.


\subsection{Constraints and sample points}
Here, we will give comments on the relevant constraints from experimental 
results. Since mass spectra and branching ratios assumed in the below studies 
are not the same as our case, following constraints cannot be directly applied 
to the gluino, squark and neutralino mass in the spectrum which we are 
considering. (For details, see each references below.)

The decay searches for $\tilde{g} \to t \bar{t} \tilde{\chi}^0_1$ via gluino 
pair production constrain the gluino-lightest neutralino mass difference 
\cite{gluinottx1ATLAS,gluinottx2ATLAS,gluinottx3ATLAS,
thirdsquarkCMS,thirdsquarkATLAS1,thirdsquarkATLAS2}. The large region of the 
simplified model with $m_{\tilde{g}} \lesssim 800$ GeV has been excluded 
\cite{thirdsquarkCMS}. 
The searches for $\tilde{g} \to bb \tilde{\chi}^0_1$
through gluino pair production also constrain the gluino and lightest neutralino 
masses \cite{gluinobbxCMS,gluinobbxATLAS,thirdsquarkATLAS2}. The simplified model 
with $m_{\tilde{g}} = 600$ GeV and $m_{\tilde{\chi}^0_1} \lesssim 400$ GeV has 
been excluded \cite{gluinobbxCMS}\footnote{
The modes of $\tilde{g} \to tb \tilde{\chi}^0_1$\cite{thirdsquarkATLAS2} 
and $\tilde{g} \to tbW \tilde{\chi}^0_1$\cite{thirdsquarkCMS} are also studied. 
}. 
These constraints would change in the case where all squark masses are nearly same 
or smaller than gluino mass because the above branching ratios decrease and 
squark-gluino production dominates the SUSY events.
Although the light sbottom is constrained also by the direct production, via the 
$\tilde{b} \to b \tilde{\chi}^0_1$ decay mode 
\cite{Aaltonen:2010dy,Abazov:2010wq,Aad:2011cw}, the sbottom mass with 
$m_{\tilde{\chi}^0_1} \gtrsim 100$ GeV has not been excluded from the searches
\footnote{
For the $\tilde{b} \to tW \tilde{\chi}^0_1$ decay mode search, see \cite{thirdsquarkCMS}.
}. 
On the other hand, the Higgsino LSP case has also been studied 
\cite{Papucci:2011wy} and stop and sbottom masses are constrained by the direct 
production search; the left-handed stop/sbottom mass, $m_{Q3} \lesssim 250$ GeV 
region has been excluded in that case. Searches of gluino decays into stop and 
sbottom also constrain the gluino and Higgsino LSP mass; their benchmark model 
with $m_{\tilde{\chi}^0_1} = 200$ GeV and $m_{\tilde{g}} \lesssim 600$ GeV has 
already been excluded (Bino LSP case has also been studied. For details, see 
\cite{Papucci:2011wy}).

For the squark direct pair production, the decay mode 
$\tilde{q} \to j \tilde{\chi}^0_1$ with $m_{\tilde{q}} \lesssim 600$ GeV and 
$m_{\tilde{\chi}^0_1} \lesssim 150 \sim 200$GeV is excluded \cite{Chatrchyan:2011ek,ATLAS:2011ad,jetsATLAS} while there is a region which is not 
almost excluded for the decay mode, $\tilde{q} \to j W \tilde{\chi}^0_1$ 
\cite{Chatrchyan:2011ek,ATLAS:2011ad,jetsATLAS}. 
The large region of the simplified 
model with $m_{\tilde{\chi}^0_1} \lesssim 200$GeV and $m_{\tilde{g}} \lesssim 700$ 
GeV are constrained by 
$\tilde{g} \to jj \tilde{\chi}^0_1$ and $\tilde{g} \to jj W \tilde{\chi}^0_1$ 
mode searches \cite{ATLAS:2011ad,jetsATLAS,jjjjWWATLASnew}. 
The CMS results exclude the simplified 
model with $m_{\tilde{g}} = 400$ GeV and $m_{\tilde{\chi}^0_1} \lesssim 350$ GeV by 
$4$ jets $+$ missing mode \cite{Chatrchyan:2011ek}. If both squark and gluino has 
been produced at the present LHC, it is also important to study the 
constraint from gluino-squark production at the LHC. In the case of 
$c_{Q_{1,2},\bar{U_{1,2}},\bar{D}} = c_{L,\bar{E}} = c_{Q_3,\bar{U}_3}$, 
large parameter space has already been excluded \cite{Chatrchyan:2011ek,ATLAS:2011ad,jetsATLAS,LeCompte:2011cn,LeCompte:2011fh}; for 
example, the simplified model with $m_{\tilde{g}} = m_{\tilde{q}} =1$ TeV has 
been excluded if the lightest neutralino mass is $0$ GeV. 
However, the small parameter region in which the mass spectrum is almost degenerate 
is not excluded; for example, the point of $m_{\tilde{g}} = 400$ GeV, 
$m_{\tilde{q}} =300$ GeV and $m_{\rm LSP} = 195$ GeV is not 
excluded\cite{jetsATLAS}.


Finally, we show the mass spectra at sample points in Tab.\ref{tab:sample point}. 
Here we assumed that $b$ term is negligibly small, for simplicity. 
At the sample point I, II and III, the lightest Higgs mass is $118$ GeV, $125$ 
GeV and $126$ GeV\footnote{
We calculate the lightest Higgs boson mass also including dominant two loop corrections\cite{Ellwanger:2009dp}.
}, and the parameters
$c_{Q_{1,2},\bar{U}_{1,2},\bar{D}} = c_{L,\bar{E}}$ is given as $1/2$, $8$ and $4$, 
respectively. 
At the sample points, Higgsinos are relatively light in order to natural EWSB. 
But, the current constraints from chargino and neutralino production can be 
satisfied \cite{Abazov:2009zi,Forrest:2009gm,Aad:2011cwa,charginolimitATLAS}. 
As shown in the previous section, there are the possibility to choice these 
discrete parameters implementing $c > 1$ in the theory. Therefore first and 
second generation squarks can be higher than gluino to avoid the current 
experimental results. Several sample points would be marginal. Then these will 
be checked at the LHC soon.

\begin{table*}[]
\begin{center}
\begin{tabular}{|c||c|c|c|c|}
\hline
Input parameters & Point I & Point II & Point III 
\\
\hline
$(c_{H_u}, c_{H_d}, c_S)$ & (0,1,0) & (0,1,0) & (0,0,1) 
\\
$(c_{q_{1,2}}, c_l)$ & (0.5, 0.5) & (8, 8) & (4, 4) 
\\
$(\lambda, \kappa, \tan\beta)$ & (0.66, 0, 2) & (0.66, 0, 2) & (0.66, 0, 3) 
\\
$\log(\Lambda/m_{\tilde{t}})$ & 4 & 4 & 1 
\\
\hline
Input parameters & value (GeV) & value (GeV) & value (GeV) 
\\
\hline
$M_0$ & 330 & 545 & 1000 
\\
$(\xi_F, C_S)$ & ($- (321)^2$, 211) & ($- (503)^2$, 253) & ($- (742)^2$, 996) 
\\
\hline
\hline
$(\mu_S,B_S)$ & (336, 103) & (487, 126) & (422, 1510) 
\\
$\mu_{\rm eff}$ & 270 & 453 & 288 
\\
\hline
\hline
particles & Mass (GeV) & Mass (GeV) & Mass (GeV) 
\\
\hline
$\tilde{g}$ & 404 & 667 & 1060 
\\
$\tilde{\chi}_{1,2}^\pm$ & 217, 378 & 408, 580 & 282, 1000 
\\
$\tilde{\chi}_5^0$ & 399 & 594 & 1000 
\\
$\tilde{\chi}_4^0$ & 354 & 516 & 984 
\\
$\tilde{\chi}_3^0$ & 312 & 503 & 452 
\\
$\tilde{\chi}_2^0$ & 290 & 466 & 304 
\\
$\tilde{\chi}_1^0$ & 191 & 380 & 262 
\\
$\tilde{u}, \tilde{c}_{L, R}$ & 322, 304 & 1590, 1580 & 2030, 2020 
\\
$\tilde{t}_{1,2}$ & 265, 384 & 424, 562 & 654, 867 
\\
$\tilde{d}, \tilde{s}_{L, R}$ & 322, 314 & 1580, 1580 & 2030, 2020 
\\
$\tilde{b}_{1,2}$ & 307, 315 & 503, 519 & 764, 774 
\\
$\tilde{e}, \tilde{\mu}_{L, R}$ & 243, 251 & 1550, 1550 & 2000, 2000 
\\
$\tilde{\tau}_{1,2}$ & 245, 255 & 1550, 1550 & 2000, 2000 
\\
$\tilde{\nu}_e, \tilde{\nu}_\mu, \tilde{\nu}_\tau$ & 243, 243, 243 & 1550, 1550, 1550 & 2000, 2000, 2000 
\\
$H^\pm$ & 480 & 795 & 319 
\\
$a_{1,2}$ & 288, 488 & 407, 799 & 316, 738 
\\
$h_{1,2,3}$ & 118, 395, 511 & 125, 621, 805 & 126, 316, 1560 
\\
\hline
\hline
mass difference & (GeV) & (GeV) & (GeV) 
\\
\hline
$m_{\tilde{g}} - m_{\tilde{\chi}_1^0}$ & 212 & 287 & 794 
\\
$m_{\tilde{t}_1} - m_{\tilde{\chi}_1^0}$ & 73.3 & 43.4 & 393 
\\
$m_{\tilde{u}_L} - m_{\tilde{\chi}_1^0}$ & 131 & 1210 & 1770 
\\
\hline
\end{tabular}
\caption{\small Mass spectrum in sample points.}
\label{tab:sample point}
\end{center}
\end{table*}

\section{Summary and discussion}

In this paper, we have shown the one of the possibility of natural 
supersymmetric spectrum by considering the NMSSM with TeV mirage mediation 
scenario. 
In this scenario,
the natural EWSB is assured by the effective low messenger scale 
$\Lambda = \mathcal{O}(1)$ TeV and proper choice of discrete parameters. 
Unlike in the MSSM, the constraint of the lightest Higgs boson mass is
satisfied also at the small $\tan\beta$ region in the NMSSM. Then the stop 
mass can be light. Furthermore, the small $\tan\beta$ can also lifts Higgsino 
mass up to the same order of those of the other SUSY particle even in the 
TeV scale mirage messenger scenario. Thus, the scenario has various mass 
spectra.

Focusing on the successful TeV scale mirage mediation, we have studied the 
possible spectrum achieving the EWSB naturally. We have shown the possible mass 
splitting in this scenario and also the nearly degenerate spectrum in which 
gluino-neutralino mass difference is around $200$ GeV. We also present the 
natural supersymmetric spectrum with the $\sim 125$ GeV 
lightest Higgs boson.
In our framework, sfermion mass hierarchy can
arise from the choice of several discrete parameter values. At some sample 
points, large squark masses with a parameter $c_{Q,\bar{U},\bar{D}_{1,2}} > 1$ 
will be viable against the experimental constraints. 
It will be important to study whether this scenario can be embedded into string 
theory or not, because the discussion in this paper depends on the modular weight 
of the matter, i.e. on how or where the matter fields are living in the bulk of 
compactification space. But it is beyond the scope of this paper.

Here, we comment on the possibility to change the parameter space, 
degenerate parameter space especially.
If we consider the TeV scale mirage mediation where the mirage condition is 
broken down in the top sector \cite{Kawagoe:2006sm}, large $m_{H_u}$ will be 
generated at the EWSB scale and spectrum would be more degenerate 
because there is a large deviation from the pure modulus mediation due to
a long running of renormalization group, 
although it is no longer an effective low messenger scale scenario. 
Moreover, if one takes larger vev of 
$T$, e.g. $T \sim 100$, $M_X \sim 10^{15}$ GeV would be obtained. In such a case, 
one may take also larger $\lambda$ at the EWSB scale and Higgs mass can be lifted 
up even in more light SUSY region while a tuning of the gauge coupling function 
depending on the gauge group will be required to realize realistic visible gauge 
coupling, which is of $\mathcal{O}(1)$. 
We will leave the study of these possibilities, and phenomenological and 
cosmological constraints on parameter space including values of 
$c_{Q,\bar{U},\bar{D},L,\bar{E}}$ to the future work \cite{future}.

We also comment on the cosmology in TeV scale mirage mediation
scenario with light and degenerate SUSY spectrum. In the parameter space 
where the correct EWSB is achieved naturally, the moduli and gravitino are 
relatively light and it is known that these light moduli and gravitino cause 
the cosmological problem \cite{Coughlan:1983ci,Banks:1993en,de Carlos:1993jw,
Nakamura:2006uc,Endo:2006zj,Kallosh:2004yh,Kawasaki:2004yh,Kawasaki:2004qu,
Kohri:2005wn,Kawasaki:2008qe}. The tachyonic squark masses at the high scale 
could also be problematic because they could lead to the unrealistic vacuum 
where $SU(3)_{\rm color}$ or $U(1)_{\rm e.m.}$ is broken. We will leave the problems because 
it can depend on the cosmological history in the universe. One of the possibility 
could be low scale inflation scenario with $H_{\inf}< m_{3/2}$ \cite{Kallosh:2004yh} 
and $T_R < 10^7$ GeV\cite{Kohri:2005wn,Kawasaki:2008qe}, and
see also the discussion about the tachyonic masses in the paper \cite{Choi:2006xb}.

To realize the correct EWSB naturally, light SUSY particles and low messenger 
scale are favored. Additionally, the current LHC experiments reports no SUSY 
signal and it can be explained by the degenerate or hierarchical SUSY spectrum 
which may be also realized by low messenger scale scenario. It is plausible that a 
(effective) low messenger scale scenario is a key to the natural EWSB avoiding 
serious fine tuning.

\subsection*{Note added:}
While this paper was being finished, new ATLAS conference note
\cite{jjjjWWATLASnew} appeared and the sensitivity for the SUSY 
signal of compressed spectrum has been really improved. This would 
play an important role to check the degenerate spectrum in our 
scenario.

\section*{Acknowledgments}

M.A. acknowledges support from the German Research Foundation (DFG) through
grant BR 3954/1-1. T.H. would like to thank Tatsuo Kobayashi for useful 
discussion.

\appendix

\section{Soft masses of the TeV scale mirage mediation}
\label{softmass_TeVmirage}


The relevant soft parameters in the Higgs sector for TeV scale mirage mediation ($\alpha = 2$) 
are approximately given by
\begin{eqnarray}
\nonumber
m_{H_u}^2(M_Z) &\simeq & 
|M_0|^2\left[
c_{H_u} -0.026 (\sum_i c_i Y_i) -
\frac{1}{4\pi^2}\left\{ -0.7+ 3|y_t|^2 + |\lambda |^2 \right\}
\log\left(\frac{\Lambda}{m_i}\right) \right] , \\
\nonumber
m_{H_d}^2 (M_Z) &\simeq & 
|M_0|^2
\bigg[ 
c_{H_u} + 0.026 (\sum_i c_i Y_i)  
- \frac{1}{4\pi^2}\left\{-0.7 + 3|y_b|^2 + |y_{\tau}|^2  + |\lambda |^2 
\right\}
\log\left(\frac{\Lambda}{m_i}\right)
\bigg] , \\
\nonumber
m_{S}^2 (M_Z) & \simeq & 
|M_0|^2\left[
c_{S} -
\frac{1}{2\pi^2}\left\{  |\lambda |^2 + |\kappa|^2 \right\}
\log\left(\frac{\Lambda}{m_i}\right)
 \right]
, \\
\nonumber
A_\lambda (M_Z) 
& \simeq &
M_0 \left[ a_{S H_u H_d} - \frac{1}{8\pi^2}\{
-1.4 + 3 |y_t|^2 + 3 |y_b|^2 + |y_{\tau }|^2 + 4 |\lambda|^2 + 2 |\kappa|^2 
\} \log\left(\frac{\Lambda}{m_i}\right)\right] , 
\\
\nonumber
A_\kappa (M_Z) 
& \simeq &
M_0 \left[a_{S^3} - \frac{3}{4\pi^2}\{|\lambda |^2 + |\kappa|^2 \}
\log\left(\frac{\Lambda}{m_i}\right)\right] .
\end{eqnarray}
This is because of $\log(\Lambda/m_{i}^2) =O(1)$ at most, where $m_i \gtrsim M_Z$ 
denotes the masses of particles in the loop diagram\footnote{
We do not take care of the difference between the masses of particles propagating in the loop
(sfermion, Higgs, Higgsino, gauginos, singlet and singlino) and energy scale at $M_Z$ in the logarithm 
since discrepancy between each logarithm will be at most of $O(1)$; they will be irrelevant to us.
}.
Furthermore, for third generation
one finds then
\begin{eqnarray}
\nonumber
A_t (M_Z) 
& \simeq &
M_0 \left[ 1 - \frac{1}{8\pi^2}\{
-8.9 + 6 |y_t|^2 + |y_b|^2  + |\lambda |^2 
\} \log\left(\frac{\Lambda}{m_i}\right)\right] , 
\\
\nonumber
A_{b} (M_Z) 
\nonumber
& \simeq &
M_0 \left[ a_{H_d Q3 \bar{D}3} - \frac{1}{8\pi^2}\{
-8.8 + |y_t|^2 + 6 |y_b|^2 + |y_{\tau}|^2  + |\lambda |^2 
\} \log\left(\frac{\Lambda}{m_i}\right)\right] , 
\\
\nonumber
A_{\tau} (M_Z) 
\nonumber
& \simeq &
M_0 \left[ a_{H_d L3 \bar{E}3} - \frac{1}{8\pi^2}\{
-1.6 + 3 |y_b|^2 + 4 |y_{\tau}|^2  + |\lambda |^2 
\} \log\left(\frac{\Lambda}{m_i}\right)\right] , \\
\nonumber
m^2_{Q3} (M_Z)
& \simeq &
|M_0|^2 \left[
c_{Q3} - 8.8\times 10^{-3}(\sum_i c_i Y_i) -\frac{1}{4\pi^2}\{
-4.4 + |y_t|^2 + |y_b|^2
\}
\log \left( 
\frac{\Lambda}{m_i}
\right)
\right] ,
\\
\nonumber
m^2_{\bar{U}3}(M_Z)
& \simeq &
|M_0|^2 \left[
c_{\bar{U}3} + 0.035(\sum_i c_i Y_i) -\frac{1}{4\pi^2}\{
-3.8 + 2|y_t|^2 
\}
\log \left( 
\frac{\Lambda}{m_i}
\right)
\right] ,
\\
\nonumber
m^2_{\bar{D}3}(M_Z)
& \simeq &
|M_0|^2 \left[
c_{\bar{D}3}  - 0.018(\sum_i c_i Y_i) -\frac{1}{4\pi^2}\{
-3.8 + 2|y_{b}|^2 
\}
\log \left( 
\frac{\Lambda}{m_i}
\right)
\right] ,
\\
\nonumber
m^2_{L3}(M_Z)
& \simeq &
|M_0|^2 \left[
c_{L3} + 0.026(\sum_i c_i Y_i) -\frac{1}{4\pi^2}\{
-0.69 + |y_{\tau}|^2 
\}
\log \left( 
\frac{\Lambda}{m_i}
\right)
\right] ,
\\
\nonumber
m^2_{\bar{E}3}(M_Z)
& \simeq &
|M_0|^2 \left[
c_{\bar{E}3} - 0.053(\sum_i c_i Y_i) -\frac{1}{4\pi^2}\{
-0.25 + 2|y_{\tau}|^2 
\}
\log \left( 
\frac{\Lambda}{m_i}
\right)
\right] .
\end{eqnarray}
Remember that $a_{ijk} = c_i + c_j + c_k = 1$ for large $y_{ijk}$.

\section{Non-perturbative $\kappa$, $\mu_S$ and tadpole}
\label{non-pert_kappa}

In this section, 
we will study the properties of the superpotential $W_{\rm NMSSM}$ in Eq.(\ref{WNMSSMmirage}).
The relevant superpotential and the corresponding soft terms is given by\footnote{
A so-called $B$-term $V \supset bH_u H_d + c.c.$ which is proportional to $\mu_S B_S$
at the quantum level would appear even though $\mu=0$ and $b=0$ at the tree level, 
if $\kappa$ and $\lambda$ are moderately large. 
}
\begin{eqnarray}
W(S, H_u , H_d) &=& \lambda S H_u H_d + \xi_F S + \frac{\mu_S}{2}S^2 + \frac{\kappa}{3}S^3 \\
V_{\rm soft} (S, H_u , H_d) 
&=& \lambda A_{\lambda} S H_u H_d + \xi_F C_S S + \frac{\mu_S}{2} B_S S^2 + \frac{\kappa}{3}A_{\kappa}S^3 + c.c..
\end{eqnarray}
In semi-realistic models within the string theories,
it is natural that there exist a lot of the exotic matter or the MSSM singlets coupled to the MSSM sector
through a compactification,
because the string theory is typically holding large gauge groups like the grand unification theory.
Such a singlet $S$ can have a charge of an anomalous $U(1)_{\rm PQ}$ symmetry,
or be a neutral modulus having shift symmetry.

Because of the symmetries mentioned above,
it can be expected that there does not exist $\kappa$, $\mu_S$ and $\xi_F$ at the perturbative level\footnote{
It is shown that a small $\kappa$ can also originate from the non-renormalizable operator 
due to the $U(1)_{\rm PQ}$ in the heterotic orbifold model \cite{Lebedev:2009ag}.
The relevant fields developing vev could be a blowing-up moduli: $W \sim \phi^n S^3 \equiv e^{-n\Phi}S^3$.
};
instantons \cite{Blumenhagen:2009qh} or gaugino condensations \cite{Dine:1985rz, Haack:2006cy}
will generate the terms\footnote{
Such a singlet $S$ could be considered as a Wilson line moduli on the D7-brane 
in the IIB orientifold models \cite{Tatar:2009jk}.
On the other hand, if D7-brane position could not be 
stabilized by closed string fluxes due to the absence of relevant ones, 
it would become light, holding larger $\kappa$ than that from non-perturbative effects.
However such a situation will not be general.
} (see discussions in the next section).
In the followings, we are assuming the case that the relevant terms are generated by such non-perturbative effects.

Let us consider the $\kappa$-coupling at the non-perturbative level:
\begin{eqnarray}
W = Be^{-8\pi^2(k_{\kappa} T + l_{\kappa}{\cal S}_0)} S^3 .
\end{eqnarray}
Then one will find $\kappa \sim (m_{3/2}/M_{\rm Pl})^{k_{\kappa}/k_h} \ll 1$
and hence $\kappa$ and $\kappa A_{\kappa}$ are negligibly small 
and the mirage mediation will not be affected by such a potential.
For a completeness, we will show $\tilde{A}_{\kappa}$ for this case:
\begin{eqnarray}
\tilde{A}_{\kappa} &=& 8\pi^2 k_{\kappa} F^T + 3s_S M_0 \\
& = & 2\frac{k_{\kappa}}{k_h}m_{3/2} + 3s_S M_0 \\
&\sim & m_{3/2} \gg |m_S|.
\end{eqnarray}

In string models, it is also natural for dimensionful terms much lower than the string scale
to be generated at the non-perturbative level.
Suppose that there is a coupling at the non-perturbative level:
\begin{eqnarray}
W & = & \frac{\mu_S}{2} S^2 \\
&=& M_{\rm Pl}Be^{-8\pi^2(k_{2} T + l_{2}{\cal S}_0)} S^2 .
\end{eqnarray}
Then for obtaining $\mu_S \simeq M_0 $ one finds
\begin{eqnarray}
k_{2} T + l_{2}{\cal S}_0 \simeq k_{h} T + l_{h}{\cal S}_0 \simeq l_0 {\cal S}_0 
.
\end{eqnarray}
Remember $8\pi^2 (k_{h} T + l_{h}{\cal S}_0) \simeq \log (M_{\rm Pl}/m_{3/2}) \simeq \log (M_{\rm Pl}/M_0)$ and 
$Ae^{-aT} \sim W_0/(aT) \sim M_0$.
Its corresponding SUSY breaking soft term $\mu_S \tilde{B}_S$ induced by modulus and SUGRA effect (compensator) is given by
\begin{eqnarray}
\tilde{B}_S 
& = & m_{3/2}\left(
2\frac{k_2}{k_h} - 1
\right) + M_0 \left( 2s_S  + 
\frac{k_v (l_0 - l_h ) + k_h l_v }{k_v (l_0-l_h)}
\right) .
\label{muBS}
\end{eqnarray}
Thus one obtains
\begin{eqnarray}
\tilde{B}_S 
& = &  M_0 \left( 2s_S  + 
\frac{k_v (l_0 - l_h ) + k_h l_v }{k_v (l_0-l_h)}
\right) \qquad
{\rm for} \qquad
\frac{k_2}{k_h}  = \frac{1}{2}.
\end{eqnarray}
Then 
\begin{eqnarray}
B_S \sim \mu_S \sim M_0
\end{eqnarray}
are found in the case that
\begin{eqnarray}
l_2 \simeq \frac{(l_0 + l_h)}{2},  \qquad \frac{k_2}{k_h}  = \frac{1}{2}.
\end{eqnarray}

Finally we shall also discuss the non-perturbative tadpole for $S$:
\begin{eqnarray}
W &=& \xi_F S \\
&=& M_{\rm Pl}^2 Be^{-8\pi^2(k_{1} T + l_{1}{\cal S}_0)} S .
\end{eqnarray}
Then for $\xi_F \simeq M^2_0$ one finds
\begin{eqnarray}
k_{1} T + l_{1}{\cal S}_0 \simeq 2(k_{h} T + l_{h}{\cal S}_0)\simeq 2 l_0 {\cal S}_0 .
\end{eqnarray}
Its corresponding SUSY breaking soft term $\xi_F \tilde{C}_S$ induced by modulus and compensator is given by
\begin{eqnarray}
\tilde{C}_S 
& = & 2 m_{3/2}\left(
\frac{k_1}{k_h} - 1
\right) + M_0 \left( s_S  + 
2 \frac{k_v (l_0 - l_h ) + k_h l_v }{k_v (l_0-l_h)}
\right) .
\end{eqnarray}
Thus the moderate $\tilde{C}_S$ would be found:
\begin{eqnarray}
\tilde{C}_S 
& = & M_0 \left( s_S  + 
2 \frac{k_v (l_0 - l_h ) + k_h l_v }{k_v (l_0-l_h)}
\right) 
\qquad
{\rm for} \qquad
\frac{k_1}{k_h}  = 1.
\end{eqnarray}
Then the appropriate relation
\begin{eqnarray}
\xi_F \sim C_S^2 \sim M^2_0
\end{eqnarray}
is realized in the case that
\begin{eqnarray}
l_1 \simeq l_0 +l_h,   \qquad \frac{k_1}{k_h}  = 1.
\end{eqnarray}

As a result, we are considering the low energy effective model
\begin{eqnarray}
\nonumber
W_{\rm NMSSM} 
&=& \lambda S H_u H_d 
+ B_1 e^{-8\pi^2 (k_h T + (l_0 + l_h){\cal S}_0)} S
+ B_2 e^{-4\pi^2 (k_h T + (l_0 + l_h){\cal S}_0)} S^2 \\
\nonumber
&& + B_3 e^{-8\pi^2 (k_{\kappa} T + l_{\kappa} {\cal S}_0)} S^3 + \dots ,\\
\nonumber
&\sim & \lambda S H_u H_d 
+ M_0^2 S + M_0 S^2 + \cdots , \\
\nonumber
&& \qquad \qquad 
\tilde{C}_S \sim \tilde{B}_S \sim M_0.
\end{eqnarray}

\subsection{RG equations with $\mu =0$}

Just below the GUT scale one obtains
\begin{eqnarray}
B_S(M_X^{-}) & \simeq & 
\tilde{B}_{S} - \gamma_{S} (M_X) m_{3/2}  \\
C_S(M_X^{-}) & \simeq & 
\tilde{C}_{S} - \frac{\gamma_{S} (M_X)}{2} m_{3/2} . 
\end{eqnarray}
Their RG equations at one-loop level are given by
\begin{eqnarray}
\frac{d}{dt}(\mu_S B_S) 
&=& \frac{1}{4\pi^2}\bigg (
\lambda^2 \mu_S ( B_S + 2 A_{\lambda}) + 2 \kappa^2 \mu_S (B_S + A_{\kappa}) 
+ 2\lambda \kappa b
\bigg )
. \\
\nonumber
\frac{d}{dt}(\xi_F C_S) 
&=& \frac{1}{8\pi^2}\bigg (
\lambda ^2  \xi_F (  C_S + 2  A_{\lambda})
+ \kappa ^2 \xi_F (  C_S + 2  A_{\kappa}) \\
&&
+ 2\lambda b (A_{\lambda} + \mu_S) + \kappa \mu_S ( B_S (A_{\kappa} + \mu_S ) + 2 m_S^2 )
\bigg ).
\end{eqnarray}
Here $b$ is usual B-term ${\cal L} \supset -b H_u H_d$ and its RG equation is given by
\begin{eqnarray}
\frac{d}{dt} b = \frac{1}{16 \pi^2} \bigg(
(3 |y_t|^2+ 3 |y_b|^2 +  |y_\tau |^2 + 6 |\lambda|^2 -g_Y^2 -3g_2^2 ) b + 2\lambda \kappa \mu_S B_S
\bigg).
\end{eqnarray}
Note that $b=0$ at the tree level; $b \sim \lambda \kappa \mu_S B_S $ will be negligibly small if $\kappa \ll 1$.
For such a case, the equations are simplified:
\begin{eqnarray}
\frac{d}{dt}(\mu_S B_S) 
&=& \frac{1}{4\pi^2}\bigg (
\lambda^2 \mu_S ( B_S + 2 A_{\lambda}) 
\bigg ) \to
\frac{d}{dt} B_S 
= \frac{\lambda^2}{2\pi^2}
 A_{\lambda}
 \\
\nonumber
\frac{d}{dt}(\xi_F C_S) 
&=& \frac{1}{8\pi^2}\bigg (
\lambda ^2  \xi_F (  C_S + 2  A_{\lambda}) 
\bigg )
\to
\frac{d}{dt}C_S 
= \frac{\lambda ^2}{4\pi^2}
    A_{\lambda}
 \\
\frac{d}{dt} b & = & 0 .
\end{eqnarray}
As a consequence, assuming that $\lambda$ originates from the similar way to the visible gauge coupling,
one then obtains 
\begin{eqnarray}
F^I \partial_I \gamma_S  = - \tilde{A}_{\lambda}\gamma_S
= -M_0 \gamma_S ,~~~{\rm where}~
\gamma_S = -\frac{\lambda^2}{4\pi^2}.
\end{eqnarray}
Therefore, in such a case with very small $\kappa$ and $b_{\rm tree} = 0$, one will again find 
the mirage soft parameters, which are similar to trilinear $A$-terms:
\begin{eqnarray}
B_S (Q) &=& \tilde{B}_S + M_0 \gamma_S \log\left( 
\frac{\Lambda}{Q}
\right) \\
C_S (Q) &=& \tilde{C}_S + \frac{M_0}{2} \gamma_S \log\left( 
\frac{\Lambda}{Q}
\right) .
\end{eqnarray}
They will be consistent with RG equations at the one-loop level.

\section{Toward the viable general NMSSM}
\label{ADS}

In this section, we will enumerate and study the possible non-perturbative effect in order to realize 
the tadpole $\xi_F$ and mass term $\mu_S$ for $S$.

First possibility is that regardless of 
whether the singlet $S$ is charged under an anomalous $U(1)_{\rm PQ}$ or not, 
stringy instantons could directly induce the relevant terms:
\begin{eqnarray}
W = A_1 e^{-8\pi^2 (k_1 T + l_1 {\cal S}_0)} S + A_2 e^{-8\pi^2 (k_2 T + l_2 {\cal S}_0)} S^2  + \cdots
\end{eqnarray}
with the appropriate $k_{1,2}$ and $l_{1,2}$ against the superpotential for the moduli stabilization, $W_{\rm KKLT} \sim 
e^{-8\pi^2 (k_h T +l_h{\cal S}_0)}$.

Second one is that if $S$ is a neutral field or modulus, one may find the instanton/gaugino condensation
via an one-loop gauge coupling \cite{Berg:2004ek, Baumann:2006th, Marchesano:2009rz} in the hidden gauge sector
\begin{eqnarray}
{\cal L}_{\rm hidden} = \int d^2 \theta \bigg( k T + l{\cal S}_0 + \frac{S^n}{8\pi^2} \bigg)
{\rm Tr}({\cal W}^{\alpha}{\cal W}_{\alpha})
\qquad (n=1 ~{\rm or}~2), 
\end{eqnarray}
for generating tadpole $\xi_F$ and $\mu_S$. Such a gauge coupling is generated
through the exchange of bulk KK modes (in the ${\cal N} = 2$ sector) coupled to the moduli
among the visible sector brane and the hidden sector one.
For such cases, via non-perturbative effects,
\begin{eqnarray}
W &\sim & \Lambda_{\rm dyn}^3 \\
& = & e^{-\frac{8\pi^2}{N}(kT + l{\cal S}_0)}e^{-\frac{S^n}{N}} \\
& = &M_0^2 + M_0^2 S + \cdots ,
\end{eqnarray}
would be obtained for $\Lambda_{\rm dyn}^3 \sim M_0^2$ and $n=1$.
Consider the case of $\Lambda_{\rm dyn}^3 \sim M_0$ and $n=2$, then
\begin{eqnarray}
W = M_0 + M_0 S^2 + \cdots 
\end{eqnarray}
would also be produced.
However, in this case, because $W \sim  M_0$ term plays a role in the KKLT moduli stabilization, 
one should obtain $k_h =k/N = k_2$; 
large $B_S$ is induced (see Eq.(\ref{muBS}))\footnote{
Similarly, when one considers the the operator of $ \int d^2 \theta H_u H_d \, {\rm Tr}({\cal W}^{\alpha}  {\cal W}_{\alpha})$ 
in the hidden gauge group, the $\mu$-term would be produced but large $b$-term $\sim m_{3/2}$ will 
also be induced due to the moduli stabilization.
}. 
Thus different mechanism should be studied for the mass term $\mu_S$ if demanded.

The third one is that
apart from the above two possibilities about $\mu_S$, 
the superpotential below for generating the moderate $\mu_S$ term and its soft parameter will be viable:
\begin{eqnarray}
W_1 
&=&\lambda_N X^{N}  +  Q_h Q_h^c ( \eta_X X + \eta_{\cal O}{\cal O})
\end{eqnarray}
with the gaugino condensation via $SU(N_c)$ gauge theory with $N_f~(< N_c)~Q_h + Q_h^c$ flavors \cite{Affleck:1983mk}:
\begin{eqnarray}
 W_{\rm ADS} & = &
(N_c -N_f) \left( 
\frac{\Lambda^{3N_c -N_f}}{{\rm det}(Q_h Q_h^c)}
\right)^{\frac{1}{N_c - N_f}} .
\end{eqnarray}
We will consider $\langle {\cal O} \rangle \ll 1$ in the Planck unit.
Here we will suppose that the couplings can depend on moduli through stringy instantons 
\cite{Blumenhagen:2009qh, Blumenhagen:2007zk}:
\begin{eqnarray}
\nonumber
\lambda_N &=& e^{-8\pi^2 (k_N T + l_N {\cal S}_0)  }   ,\quad
\eta_{X}= e^{-8\pi^2 (k_{X} T + l_{X}{\cal S}_0)} , \quad 
\eta_{\cal O} = e^{-8\pi^2 (k_{{\cal O}} T + l_{{\cal O}}{\cal S}_0)}  , \\
\Lambda^{3N_c - N_f} 
&=& e^{-8\pi^2  (k_c T + l_c {\cal S}_0) } .
\end{eqnarray}
For instance, though a case that $\eta_{\cal O}$ is the constant $(k_{\cal O}=0,~l_{\cal O}=0)$ will be possible,
we just generalized the expressions.
Thus after integrating out mesons $(Q_h Q_h^c)$ and $X$ with $W = W_1 + W_{\rm ADS}$, 
the low energy effective superpotential is given by
\begin{eqnarray}
\nonumber
W_{\rm eff} &=& 
 \sum_{p=0} B_p  \eta_{\cal O}^p
(\eta_X^{-N} \lambda_N)^{\frac{ p N_c - N_f}{N N_c - N_f}}
\left(\Lambda^{3 N_c - N_f}\right)^{\frac{ (N - p)}{ N N_c - N_f }}  {\cal O}^p
\\
&=&
\sum_{p=0} B_p 
\exp \left[- 8\pi^2 
( K_p T + L_p {\cal S}_0 )
 \right] {\cal O}^p \equiv \sum_{p=0} \Omega_p {\cal O}^p .
\end{eqnarray}

Below we will consider two cases of ${\cal O}=S$ and ${\cal O} =S^2$ 
for producing the tadpole $\xi_F$ and the mass term $\mu_S$.

\begin{itemize}

\item ${\cal O} = S$

In order to obtain appropriate $\xi_F$, $\mu_S$ and their soft terms simultaneously,
one might set 
\begin{eqnarray}
{\cal O} = S  .
\end{eqnarray}
Through the equations
\begin{eqnarray}
K_1 = k_h = 2K_2, \qquad L_1 = l_0 + l_h = 2 L_2 ,
\end{eqnarray}
one would find
the appropriate relation
\begin{eqnarray}
\Omega_0 \sim M_0^3 , \quad
\xi_F \sim C_S^2 \sim M_0^2 ,\quad
\mu_S \sim B_S \sim M_0, \quad
\kappa \sim  \Omega_3 \sim 1 .
\end{eqnarray}
However this means
\begin{eqnarray}
W_{\rm eff} 
&=&
M_0^3 \sum_{p=0} B_p 
 \frac{S^p}{M_0^p} .
\end{eqnarray}
Thus, it is impossible to use Taylor expansion on $S$ in the regime $S \gtrsim M_0$.
Furthermore, it would be implausible to discuss integration of massive modes 
since it is expected that their mass scales are given by about $M_0$, which is the soft SUSY breaking mass scale.
Hence this case will not interest us. 

\item ${\cal O} = S^2$

For the mass term $\mu_S$, set 
\begin{eqnarray}
{\cal O} = S^2  , \qquad
2K_1 = k_h , \qquad 2 L_1 = l_0 + l_h .
\end{eqnarray}
Then, the constant $\Omega_0$ is given by
\begin{eqnarray}
\nonumber
\Lambda_S^2 & \equiv & \frac{\Omega_0}{M_0} 
, \qquad M_0 = e^{-4\pi^2 (k_h T + (l_0 + l_h){\cal S}_0)} ,
\\
\nonumber
& = & M_{\rm Pl}^2  \exp\bigg[
-\frac{8\pi^2 {\cal S}_0}{N-1} \left( 
\frac{l_0 + l_h}{2}
+
 N (l_X - l_{\cal O})  - l_N 
\right)
\bigg]  
\\
\nonumber
&& 
\times \exp\bigg[
-\frac{8\pi^2 T}{N-1} \left( 
\frac{k_h}{2} 
+  N (k_X -k_{\cal O})  - k_N  \right)
\bigg] .
\end{eqnarray}
Thus, as far as the conditions that
\begin{eqnarray}
N (l_X - l_{\cal O})  - l_N  \geq 0
, \qquad
(k_X -k_{\cal O}) N - k_N  \geq 0
\end{eqnarray}
are satisfied, the KKLT stabilization will be viable, i.e. $\Omega_0 \ll M_0$, and
it will be possible to obtain
\begin{eqnarray}
\Omega_1 \sim \mu_S \sim B_S \sim M_0.
\end{eqnarray}
Then one will find
\begin{eqnarray}
W_{\rm eff} &=& \Omega_0 \sum_{p=0}  \left(\frac{\Omega_0}{M_0}\right)^{-p} {\cal O}^p \\
& = & M_0 \Lambda_S^2 \sum_{p=0} \left(\frac{S}{\Lambda _S}\right)^{2p} ,\qquad \Lambda_S \equiv \sqrt{\frac{\Omega_0}{M_0}} \gg M_0 
\end{eqnarray}
for $\Omega_0 \gg M_0^3$.
In special, for the case that
\begin{eqnarray}
l_N = (l_X -l_{\cal O})N - \frac{(l_0 + l_h)}{2} (N-2), \quad
k_N = (k_X -k_{\cal O}) N  - \frac{k_h}{2}(N - 2),
\end{eqnarray}
$\Lambda_S^2 = M_0$ is realized. This case would be desirable one for the tadpole $\xi_F$
when the hidden sector gauge coupling in the ADS superpotential has the $S$ dependence:
\begin{eqnarray}
f_c = k_c T + l_c {\cal S}_0 + \frac{S}{8\pi^2}.
\end{eqnarray}
This is because the ADS superpotential is then deformed as
\begin{eqnarray}
W_{\rm ADS} &\to & 
W_{\rm ADS}|_{S=0} + S \partial_S W_{\rm ADS}|_{S=0} + \cdots \\ 
&=& M_0^2 + M_0^2 S + M_0^2 S^2 +\cdots .
\end{eqnarray}

On the other hand, using $\Lambda_S$ and $M_0$, the relevant vevs of heavy modes are written by
\begin{eqnarray}
\nonumber
\eta_{\cal O} \langle (Q_h Q_h^c) \rangle &\sim & {M_0}
,  \qquad
\langle X \rangle \sim 
\frac{\eta_{\cal O}}{\eta_X} \Lambda_S^2 
\sim 
\bigg(
\frac{\eta_X}{\eta_{\cal O} \lambda_N}
M_0
\bigg)^{\frac{1}{(N-1)}}
.
\end{eqnarray}

Then, the viable superpotential will be summarized:
\begin{eqnarray}
W &=& \lambda_N X^N + Q_h Q_h^c (\eta_X X + \eta_{\cal O}S^2 )  + W_{\rm ADS} + W_{\rm tadpole}.
\end{eqnarray}
Here
\begin{eqnarray}
\nonumber
W_{\rm tadpole} &=&  
\eta_{\cal O}^2(Q_h Q_h^c)^2 S  \qquad
  \\
\nonumber
 & {\rm or} & \frac{\lambda_N^2\eta_{\cal O}^2}{\eta_X^2} X^{2(N-1)} S  \qquad
  \\
\nonumber
  &{\rm or}& \frac{\lambda_N\eta_{\cal O}^2}{\eta_X}(Q_h Q_h^c) X^{(N-1)} S  \qquad
 \\
\nonumber
  &{\rm or}& S \partial_S W_{\rm ADS} 
 \\
\nonumber
  &{\rm or}& A_1 e^{-8\pi^2(k_h T + (l_0 + l_h ){\cal S}_0)} S.
\end{eqnarray}
In the last line, we included the possibility from an (stringy) 
instanton/gaugino condensation we considered above.
Finally if one
sets ${\cal O}= H_u H_d$ with the ADS superpotential, 
the moderate $\mu$-term would be generated similarly \cite{Casas:1992mk, Choi:2006xb}.

\end{itemize}

More plausible discussion with an $U(1)_{\rm PQ}$ symmetry \cite{Jeong:2011jk}\footnote{
If there are additional moduli, such a global symmetry will be viable.
For instance, consider the K\"ahler potential
$K_{\rm moduli} = -3 \log(T'+\bar{T}' - (G+\bar{G})^2/{\cal S}+\bar{\cal S})$.
In type IIB orientifold model, this $G$-modulus belongs to $h^{1,1}_-$ 
one and necessarily charged under anomalous $U(1)$ \cite{Grimm:2004uq};
a linear combination of $T'$ and $G$ will be absorbed into the $U(1)$ multiplet 
and then gets heavy if $T'$ is also charged \cite{Higaki:2011me}.
The remaining modulus is a neutral modulus denoted by $T$.
Hence there can be a neutral instanton effect $W \sim e^{-aT}$ \cite{Grimm:2011dj}
and $T$ consequently plays the role in the SUSY-breaking modulus on which we are focusing.
And then a global $U(1)_{\rm PQ}$ symmetry emerges. 
This symmetry is to be broken down to a discrete symmetry by non-perturbative effect 
\cite{Banks:2010zn, BerasaluceGonzalez:2011wy}.
} or $R$-symmetry \cite{Araki:2008ek}
will be very interesting from the view point of model-building. We will leave it as a future work.

\end{document}